\documentclass[twocolumn,showpacs,preprintnumbers,amsmath,amssymb]{revtex4}

\usepackage{graphicx}
\usepackage{dcolumn}
\usepackage{bm}

\begin{document}


\title{Gundlach oscillations and Coulomb blockade of Co nano-islands on MgO/Mo(100) investigated by scanning tunneling spectroscopy at 300 K}

\author{C. Pauly}
\author{M. Grob}
\author{M. Pezzotta}
\author{M. Pratzer}
\email{pratzer@physik.rwth-aachen.de}
\author{M. Morgenstern}
\altaffiliation[]{2nd Institute of Physics B , RWTH Aachen University, 52056 Aachen, Germany; J\"ulich-Aachen Research Alliance: Fundamentals of
Future Information Technology (JARA-FIT)}


\date{\today}

\begin{abstract}

Ultrathin MgO films on Mo(100) with a thickness up to 12\,ML are studied by scanning tunneling microscopy and spectroscopy at room temperature. The spatial variation of the work function within the MgO film is mapped by field emission resonance states (Gundlach oscillations) using $dz/dU$ spectroscopy. We found circular spots with significantly reduced work function ($\Delta \Phi=0.6$\,eV), which are assigned to charged defects within the MgO film. On top of the MgO films, small Co cluster are deposited with an average contact area of $A_{\rm Co}\simeq 4$\,nm$^2$. These islands exhibit Coulomb oscillations in $dI/dU$-spectra at room temperature. Good agreement with orthodox theory is achieved showing variations of the background charge $Q_0$ for islands at different positions, which are in accordance
with the work function differences determined by the Gundlach oscillations.

\end{abstract}

\maketitle

\section{Introduction}

Magnesium oxide (MgO) is a preferred insulator for magnetic tunnel junctions \cite{Moodera,Parkin1,Parkin2,Ikeda08,Meyerheim,Heinrich,Pacchioni,Wulfhekel,Buhrman} and is extensively used as a template for the microscopic study of catalytic reactions \cite{Moia,Pfnur,Schneider,Heyde,Freund2,Freund4,Freund5,Gallagher,Nilius,Freund,Freund3,Renaud}. Both fields rely on the good epitaxial quality of MgO films, while the former was additionally boosted by the theoretical insight of a symmetry induced strong spin selectivity of MgO(001) \cite{Mathon01}.MgO might also be an excellent template for nanoelectronic studies by scanning probe microscopy similar to the ones that have been performed recently on NaCl with respect to charge manipulation and bond formation \cite{Repp1,Repp2,Repp3} or on Al$_2$O$_3$ and CuN with respect to the determination of magnetic properties of individual atoms on a substrate \cite{Heinrich1,Heinrich2,Heinrich3}.\\
Thin films of MgO, exhibiting a wide band gap and a simple rock salt structure, grow epitaxially on different metal substrates as Ag(001) \cite{Moia,Pfnur,Schneider,Heyde,Freund2,Freund4,Freund5}, Fe(001) \cite{Meyerheim,Heinrich,Pacchioni,Wulfhekel,Buhrman} and Mo(001) \cite{Gallagher,Nilius,Freund,Freund3,Renaud}. Here, we choose Mo, since it allows high annealing temperatures exceeding 1000\,K, which might foster an improved MgO film quality. The remaining major defects are color centers exhibiting defect states within the band gap with energies depending on the defect position \cite{Freund2,Freund4,Freund5}. Work function differences between different layer thicknesses of MgO have been determined by field emission resonance (FER) spectroscopy and photon mapping to be up to 650\,meV \cite{Freund3}.\\
In this work, we employ scanning tunneling microscopy (STM) and spectroscopy (STS) in order to investigate pure MgO films grown epitaxially on Mo(100) as well as Co nano-islands deposited on top of the MgO.  We observe an increase of the MgO band gap $E_{\rm Gap}$ with increasing film thickness which reached $E_{\rm Gap}=7.1$\,eV at a thickness of 11 monolayers (ML). Mapping the work function $\Phi (x,y)$ by the energetic shift of the first Gundlach oscillation reveals a strong spatial variation up to about $\Delta \Phi = 0.6$\,eV. This shift could be correlated with a peak within the gap known to be caused by charged color centers \cite{Freund2}. This explains the work function shifts straightforwardly as caused by color centers located within the topmost two MgO layers.\\
Probing Co islands on top of the MgO film, we observe Coulomb oscillations by $dI/dU$ spectroscopy with energies in excellent agreement with orthodox theory \cite{Likharev}. A varying Coulomb gap around $U=0$\,V  indicates a fluctuating background charge, which could also be attributed to the charged color centers close to the surface. The width of the Coulomb peaks of $0.3-0.4$\,eV is much larger than the energy resolution of the experiment. This is probably due to a large coupling of the cluster charge to the phonons of the ionic insulator \cite{Repp4}. Notice that Coulomb oscillations of metal clusters on insulating thin films have rarely been observed at room temperature so far  \cite{Donkersloot, Wulfhekel, Hoffmann}.

\section{Experiment}

The experiments were carried out using an ultrahigh vacuum system (base pressure $p=8\cdot 10^{-11}$\,mbar) consisting of two separately pumped chambers. The main chamber is equipped with a 4-grid low-energy electron diffraction (LEED) optics, an electron beam heating stage and a modified Omicron STM for measurements at room temperature. A side chamber is equipped with the facilities for MgO growth, namely an electron beam induced Mg evaporator, a leak valve for oxygen and a heating stage positioned directly in front of the evaporator. Prior to MgO preparation, the Mo(100) crystal was cleaned by cycles of annealing in an oxygen atmosphere ($p_{\rm O_2}=5\cdot 10^{-7}$\,mbar) at 1400\,K followed by subsequent flashing to 2300\,K. MgO films were prepared by molecular beam epitaxy of magnesium in an $O_2$ environment at a partial pressure of $p_{\rm O_2}=1.5\cdot 10^{-7}$\,mbar keeping the substrate at room temperature. The deposition rate of $0.5$\,ML/min has been controlled by means of a quartz microbalance operating at $10$\,MHz. After MgO deposition, the samples were annealed at 1100\,K for 10\,min. We checked the purity of the substrate and the MgO films by means of Auger electron spectroscopy and LEED. All STM and STS measurements were carried out with an electrochemically etched tungsten tip with the bias applied to the sample. STS data ($dI/dU$, $dz/dU$) were acquired by lock-in-technique applying a modulation voltage with amplitude $\sqrt{2}\cdot U_{\rm mod}$. $dI/dU$ curves are recorded with feedback off after stabilizing the tip at voltage $U_{\rm stab}$ and current $I_{\rm stab}$.

\section{Results and Discussion}

\begin{figure}[htb!]
\includegraphics[width=8.1cm]{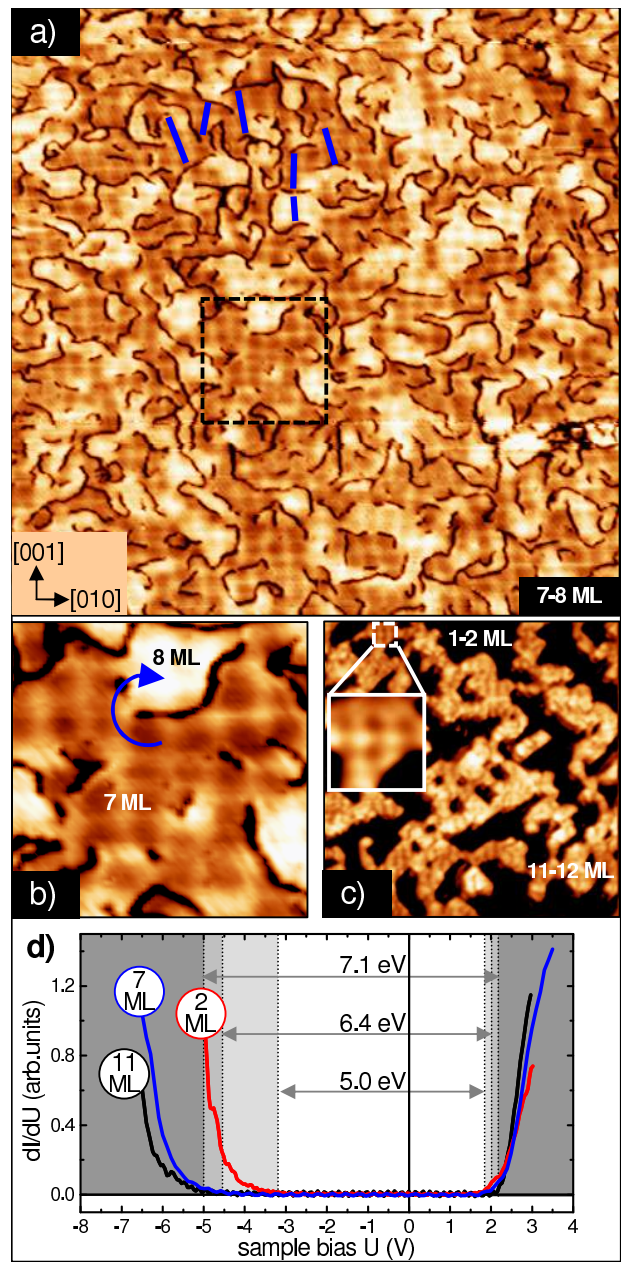}
\caption{\label{morphology}(Color online) a) ($200\times200$)\,nm$^2$ STM image of 7-8\,ML MgO film on Mo(100) ($U=3$\,V, $I=0.5$\,nA); the typical Moir\'e pattern caused by the lattice mismatch of Mo and MgO is visible; the marked crystallographic directions are determined by LEED. The lines emphasize the deviation of the Moir\'e pattern from the Mo[001] direction within different MgO facets. b) Zoom into the MgO film as marked by a dashed line in image a); the Moir\'e pattern and a screw dislocation indicated by an arrow is visible. c) MgO sample consisting of 11-12\,ML thick islands and valleys in between consisting of 1-2\,ML MgO. The inset shows the Moir\'e pattern still appearing on MgO islands of 11\,ML. d) Spectra of $dI/dU (U)$ measured on MgO areas of 2\,ML, 7\,ML and 11\,ML thickness as indicated ($U_{\rm stab}=3$\,V, $I_{\rm stab}=0.45$\,nA, $U_{\rm mod}=40$\,mV); the deduced band gap is marked.}
\end{figure}

Figure \ref{morphology} a) shows an STM image of an MgO film on Mo(100) with a thickness of 7-8\,ML. The nearly completely covered surface exhibits 1.3\,nm deep grooves (black lines) which mainly separate differently tilted MgO facets. A Moir\'e pattern caused by the 5.4\,\% lattice-misfit of Mo and MgO is clearly visible all over the film. Previous LEED studies indicated this superstructure to exist up to a MgO thickness of 12\,ML \cite{Renaud}, but it has never been observed for such thick films directly by STM. In addition, we observed large variations in the periodicity of the Moir\'e pattern ranging from 3\,nm to 8.3\,nm compared to the calculated value of 5.35\,nm. Further different MgO facets show tilting angles with a deviation of up to 20$^\circ$ from the Mo[001] direction as indicated by the lines in Figure \ref{morphology} a). This may be caused by relaxation effects during cooling down the sample from 1100\,K. Figure \ref{morphology} b) displays a zoom into the film highlighting that the grooves sometimes appear at step edges originating from screw dislocations within the film.\\
Figure \ref{morphology} c) shows another MgO film, nominally prepared under the same conditions as the film shown in a) and b). Here, we do not find a smooth film, but large valley areas covered by only 1-2\,ML MgO and, in between, higher MgO islands with thicknesses up to 11-12\,ML. The thickness of the MgO within the valleys is consistently determined by observation of step heights within the valleys up to 2 ML and comparison of the total MgO film topography with the nominally deposited amount of Mg. The origin of this different growth type is not known, but may be caused by a small amount of molybdenum oxide remaining on the substrate prior to MgO preparation, which might alter the MgO nucleation. However, we could not detect the corresponding oxygen on the cleaned substrate by Auger electron spectroscopy. The Moir\'e superstructure induced by misfit can still be observed on MgO islands with a thickness of 11\,ML (see inset of figure \ref{morphology} c).\\

\begin{figure*}[htb!]
\includegraphics[width=17cm]{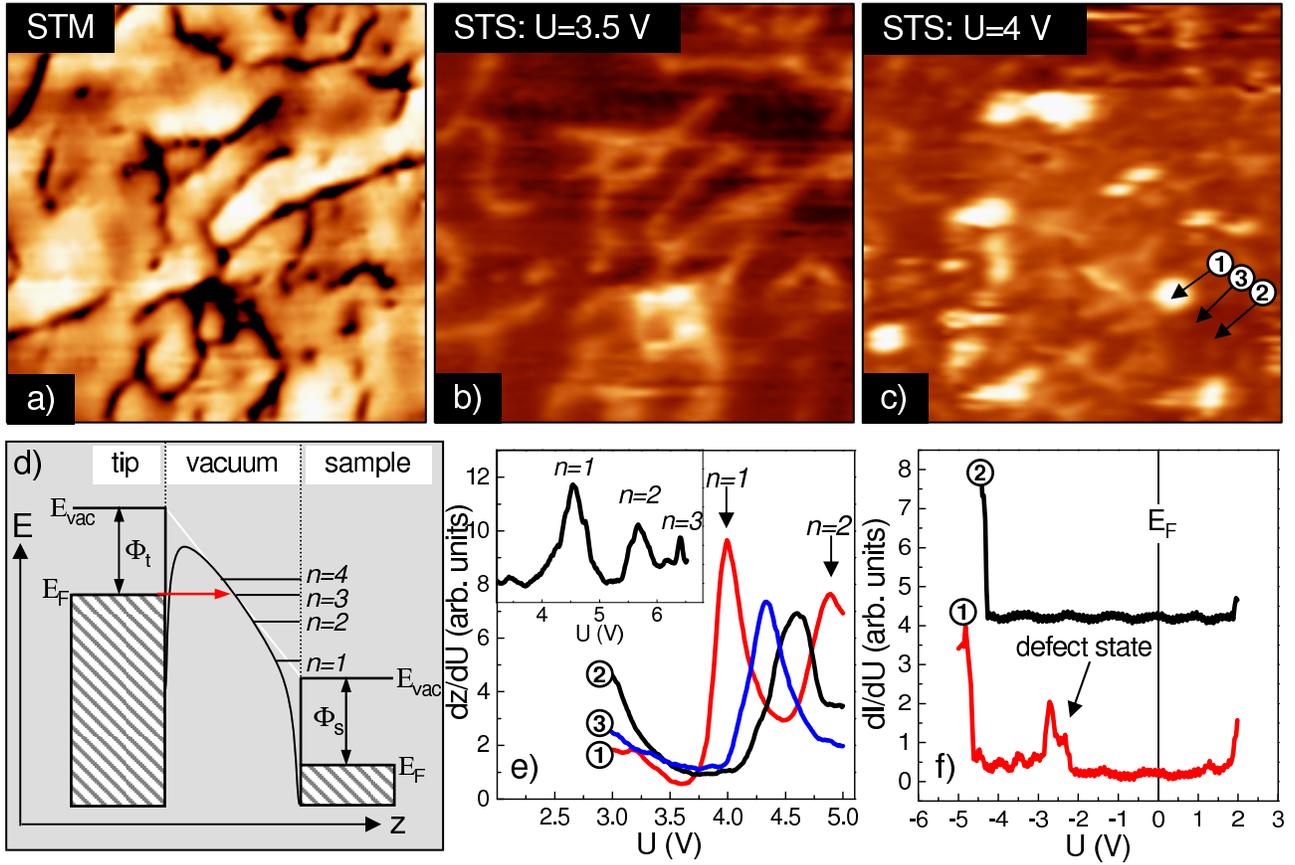}
\caption{\label{spectroscopy}(Color online) a) ($40\times40$)\,nm$^2$ STM image of 7-8\,ML thick MgO film ($U=3.5$\,V, $I=0.5$\,nA). b) $dz/dU$-map recorded at $U=3.5$\,V of the same area as shown in a) ($I=0.5$\,nA, $U_{\rm mod}=60$\,mV); only the grooves in between the islands exhibit spectroscopic contrast. c) $dz/dU$ map recorded at $U=4.0$\,V of the same sample area as a) and b) ($I=0.5$\,nA, $U_{\rm mod}=60$\,mV); additional bright spots appear on parts of the MgO film. d) Sketch of the potential scheme within the tunneling barrier explaining the field resonance states; $E_F$, $E_V$ and $\Phi$ marked at tip and sample are the Fermi levels, vacuum levels and work functions, respectively; the black line in the vacuum region is the vacuum level influenced by applied electric field and the image potential of tip and sample (white line without image potential); FER states are marked by $n=1$, $n=2$, $n=3$ and $n=4$. e) $dz/dU$-spectra taken at the positions marked in c) ($U_{\rm stab} = 3.5$\,V, $I_{\rm stab} = 0.5$\,nA, $U_{\rm mod}=100$\,mV); the first two FER states $n=1$ and $n=2$ are labeled for curve 1; the $n=1$ state shifts by about $\Delta E=620$\,meV between curve 1 and curve 2. Inset: three FER peaks measured on a dark region of c). f) $dI/dU (U)$-spectra recorded on two of the positions marked in c) ($U_{\rm stab} = 4.0$\,V, $I_{\rm stab} = 0.5$\,nA, $U_{\rm mod} = 26$\,mV); the apparent band gap exhibits a peak at $U=-2.7$\,V only in curve 1, which indicates the defect state.}
\end{figure*}

The band gap of the MgO film is measured by STS as shown in figure \ref{morphology} d). The edge of the band gap is determined by the point where positive $dI/dU$ intensity is larger than the noise level.
 We checked that this method is sufficiently reliable by recording $dI/dU$ curves at different stabilization parameters.
 An increase of the band gap from 5\,eV on 2\,ML MgO (measured within the valley area) towards 7.1\,eV on 11\,ML MgO (the high islands) can be deduced approaching the bulk value of 7.8\,eV \cite{Roesler}. Notice that the gap grows asymetrically indicating that electron affinity of MgO is less affected by film thickness than ionization energy.\\
Electronic properties of the smooth film are further investigated by STS. Field emission resonance states (FER), i.e. Gundlach oscillations, are routinely observed as shown, e.g., in the inset of figure \ref{spectroscopy} e) exhibiting peaks at voltages of 4.5\,V, 5.7\,V and 6.4\,V. Such FER states are known to appear if the bias voltage exceeds the work function of the sample. The origin of these states is sketched in figure \ref{spectroscopy} d). They are tunneling resonances caused by the constructive interference of electron waves, which are reflected back and forth between the potential step at the sample surface and the potential wall given by the vacuum level.  Due to the high electric field at the required voltages to probe FER states by $dI/dU(U)$, tip changes are frequent. To avoid that, we measured $dz/dU$-spectra with a closed feed-back loop keeping the tunneling current constant. This retracts the tip automatically at higher voltage, thereby limiting the electric field. We used lock-in technique to determine the $dz/dU$ signal. Figure \ref{spectroscopy} a) shows a constant current image of a 7\,ML thick MgO film. The corresponding $dz/dU$ image taken at 3.5\,V (figure \ref{spectroscopy} b) shows spectroscopic contrast only at the deep grooves (bright lines). Figure \ref{spectroscopy} c) displays the $dz/dU$ signal of the same area taken at a bias voltage of 4\,V, where bright spots randomly distributed over the MgO film appear additionally to the fading lines originating from the grooves. The reason for the spectroscopic contrast can be understood by recording single point $dz/dU$ spectra on bright and dark areas. Figure \ref{spectroscopy} e) shows three spectra mapping the transition from a bright spot to the dark surrounding. The first FER state shifts continuously from $U(n=1)=4$\,V (spectrum 1), measured in the center of the bright spot, over $U(n=1)=4.3 $\,V (spectrum 3) to $U(n=1)=4.6$\,V (spectrum 2), measured on the dark surrounding 10\,nm away from the center of the bright spot. The maximum shift in peak energy of $\Delta E_G=600$\,meV indicates a large change in the MgO work function. Note, that the energy shift of the first FER peak gives only an approximation of the work function difference because of the additional image potential, which reduces the FER energies by a value which is stronger at lower FER energy. Therefore, the peak-shift is only a lower limit of the real change of the work function.\\
Furthermore, $dI/dU$ spectroscopy of the band gap region is measured on bright spots and dark parts of the $dz/dU$ image, as exemplified in figure \ref{spectroscopy} f). The $dI/dU$ curves show an additional peak at about $-2.7$\,V only on the bright spots. Referring to theoretical calculations this occupied state can be attributed either  to charged F$^+$ color centers or to MgO surface defects at 3-fold coordinated surface sites \cite{Shluger}. The defect state around $-2.7$\,V has been previously observed at MgO-islands on Ag(001) \cite{Freund2, Freund5}, but in contrast to one of these works, we do not observe an empty state at $\approx$$+1$\,V. This might be due to a shift of the empty state towards the conduction band of the MgO within the MgO/Mo(100) system.
It might be surprising that the defects mapped by FER spectroscopy appear relatively large, i.e. about $3-4$\,nm in diameter. This size is definitely larger than the defect state mapped at $-2.7$\,V in previous low temperature STM measurements \cite{Freund2}. The major difference between the two methods is that  we detect the electric potential of the charge and not the state of the defect.
Estimating the potential of a point charge about 1.5 nm away from the center still results in a potential shift of 200\,meV.
Together with our energy resolution of $\Delta E=170$\,meV in Fig. \ref{spectroscopy} c), this already justifies the observed
size of the charge within FER images. However, the field emission process is more subtle depending on details of the
3D potential between tip and substrate \cite{Lang, Gadzuk} and more detailed calculations are required for a more quantitative analysis of the apparent size of a charged defect in FER images. Of course charged defects of the size of a few atoms cannot be excluded.\\
The depth of the singly positive charge with respect to the MgO surface ($d_d$) can be estimated by the Coulomb potential of a point charge  ($e\cdot U_c$, $e$: elementary charge) leading to $d_d=e/(4\pi\varepsilon\varepsilon_0 U_c)=0.25$\,nm ($\varepsilon_0$: dielectric constant of vacum), if the bulk value of the dielectric constant of MgO $\varepsilon =9.8$ \cite{Schuele} is used and $d_d=0.5$\,nm if half the value of $\varepsilon$ is used as a lower estimate considering the presence of the surface. This indicates that the corresponding defect must be located very close to the MgO surface, probably in the subsurface layer of the insulating film. Notice, that previous FER spectroscopy and photon maps of MgO films on Mo(001) showed energy shifts in the same range (650\,meV), but between thick and thin MgO-islands \cite{Freund3} and not as spatial fluctuations within a smooth film. \\

\begin{figure}[htb!]
\includegraphics[width=8.5cm]{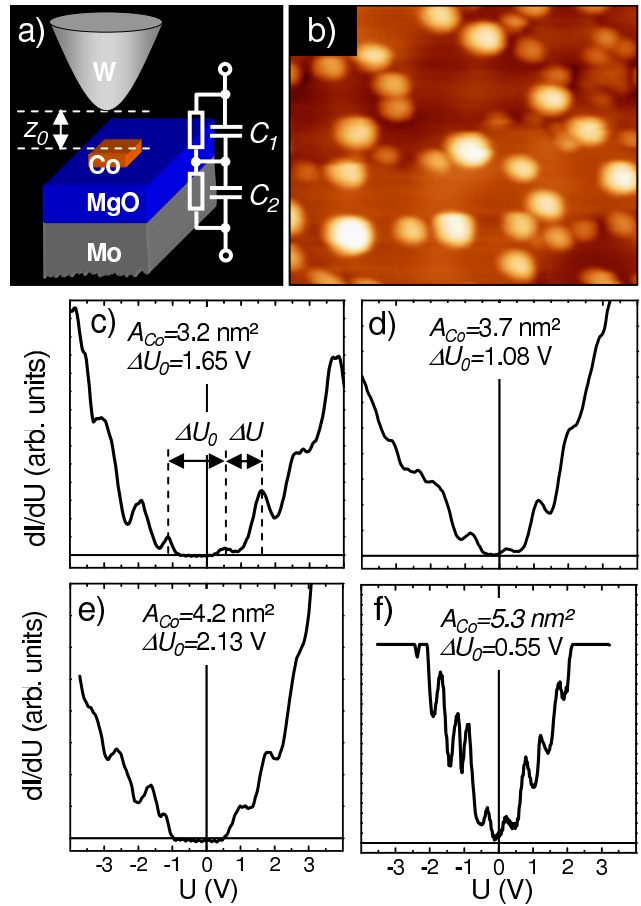}
\caption{\label{blockade}(Color online) a) Schematic drawing of the tunneling junction consisting of tip/vacuum/Co-island and Co-island/MgO/Mo(100) together with equivalent circuit diagram. b) STM image ($25\times 20$)\,nm$^2$ of Co nano-islands on MgO ($U=3.5$\,V, $I=0.5$\,nA). c)-f) $dI/dU(U)$-spectra recorded on Co islands with different sizes $A_{\rm Co}$ as indicated ($U_{\rm stab}=3.5$\,V, $I_{\rm stab}=0.7$\,nA, $U_{\rm mod}=40$\,mV); the Coulomb gap $\Delta U_0$ and the charging voltage $\Delta U$ are marked in c); spectrum f) is measured on a Co island on 10\,ML MgO, spectra c)-e) are recorded on an island of 11\,ML MgO.}
\end{figure}

Next, we prepared Co nano-islands on top of the MgO film by molecular beam epitaxy at room temperature. As illustrated in figure \ref{blockade} b), Co forms compact islands exhibiting an average area of $A_{\rm Co}\simeq 4$\,nm$^2$ and a height of 2-4\,ML. The Co nano-islands are well separated from each other. This is an ideal system for Coulomb blockade, since it consists of two tunneling junctions given by W-tip/vacuum/Co-island and Co-island/MgO/Mo(100). The equivalent electrical circuit is drawn in figure \ref{blockade} a) consisting of two capacitors ($C_1$, $C_2$) and two resistors ($R_1$, $R_2$) in parallel. Single electron tunneling occurs, if the charging energy per electron is much larger than the thermal energy, i.e.

\begin{equation}\label{single}
\frac{e^2}{C_1+C_2}>>kT.
\end{equation}

Thus, the capacity must be as small as a few atto-Farad in order to observe Coulomb staircases in the $I(U)$-signal, respectively, peaks in the differential conductivity $dI/dU(U)$ at $T=300$ K. Moreover, sequential electron tunneling requires resistances of the two junctions higher than the resistance quantum $h/2e^2 = 12.9$ k$\Omega$, which, however, is easily achieved within our setup. Figure \ref{blockade} c)-f) show $dI/dU(U)$-spectra for several Co islands with different sizes. Within the band gap of MgO (from $-5$\,V to $+2$\,V), several peaks are visible indicating Coulomb staircases. The central Coulomb gap $\Delta U_0$ and the distance between adjacent peaks at higher energy $\Delta U$ is marked in Fig. \ref{blockade}c). The resulting values of $\Delta U_0$ are given in Fig. \ref{blockade}c)-f) together with the contact area $A_{\rm Co}$ of the corresponding island.\\
To compare the measured $\Delta U$ with orthodox theory \cite{Likharev}, we calculate the expected $\Delta U^\star=e/(C_1+C_2)$ \cite{Donkersloot,Tinkham}. The capacities $C_1$ and $C_2$ of the two tunneling junctions are regarded as two plate capacitors with the area of the Co nano-island $A_{\rm Co}$ resulting in $C=\varepsilon\varepsilon_0 A_{\rm Co}/d$. Thus, $C_2$ can be determined using the measured thickness of the MgO-film as $d$ and the dielectric constant of MgO $\varepsilon_{\rm MgO}=9.8$ \cite{Schuele}. For calculation of $C_1$, the absolute tip-sample distance $z_0$ must be known. This number is difficult to determine directly by STM. We estimate $z_0$ by extrapolating $I(z)$ curves to the distance, where the conductance would reach the conductance quantum $G_0=2e^2/h$. This point is regarded as the point of mechanical contact corresponding to $d=0$ nm \cite{Berndt,Mashoff}. This results in a tip-surface distance $z_0$ at $I_{\rm stab}$ and $U_{\rm stab}$ of:

\begin{equation}
z_0=-\frac{1}{2\kappa}\ln\left\lbrace\frac{1}{G_0}\frac{I_{\rm stab}}{U_{\rm stab}}\right\rbrace.
\end{equation}

The decay constant $\kappa$ for a planar tunneling junction is given by \cite{Ukraintsev}:

\begin{equation}\label{kappa}
\kappa=\sqrt{\frac{m_{\rm e}}{\hbar^2}\left(\Phi_{\rm W}+\Phi_{\rm Co}-\left|eU_{\rm stab}\right|\right)}.
\end{equation}

We used the work function of the densely packed W(110) surface $\Phi_{W}=5.25$\,eV for the W tip and a Co work function of $\Phi_{Co}=5$\,eV \cite{Michaelson}. This leads to $z_0\simeq 6.7$ \AA, which is taken as $d$ for $C_2$. Table \ref{results} shows the calculated values $C_1$, $C_2$ and $\Delta U^\star$ in comparison with the measured $\Delta U$ for different island sizes. Excellent agreement between $\Delta U$ and $\Delta U^\star$ is found without any fit parameter.\\
The Coulomb gap $\Delta U_0$ depends, in addition, on the background charge $Q_0$ and is given by \cite{Likharev, Tinkham}:

\begin{table}
\caption{\label{results} Table of calculated values for the serial capacities $C_1$, $C_2$ and the Coulomb peak distances $\Delta U^\star$ compared with the measured values $\Delta U$ as taken from figure \ref{blockade}. In addition, the lateral size of the Co island, the determined background charge $Q_0$, the corresponding local change of Co work function $\Delta \Phi_{\rm Co}$
and the average full width at half maximum of the Coulomb peaks $\Delta E_{\rm p}$ is given.}
\begin{ruledtabular}
\begin{tabular}{cccccccc}
\textbf{size} & \textbf{$C_1$} & \textbf{$C_2$} & \textbf{$\Delta U$} & \textbf{$\Delta U^\star$} & \textbf{$|Q_0|/e$} & \textbf{$\Delta \Phi_{\rm Co}$} & \textbf{$\Delta E_{\rm p}$}\\
3.2\,nm$^2$ & 41\,zF & 120\,zF & 1.03\,V & 1.00\,V & 0.19 & 385\,meV & 395\,meV\\
3.7\,nm$^2$ & 47\,zF & 139\,zF & 0.90\,V & 0.86\,V & 0.26 & 453\,meV & 380\,meV\\
4.2\,nm$^2$ & 53\,zF & 157\,zF & 0.84\,V & 0.76\,V & 0 & 0\,meV & 385\,meV\\
5.3\,nm$^2$ & 67\,zF & 219\,zF & 0.50\,V & 0.56\,V & 0.34 & 408\,meV & 290\,meV\\
\end{tabular}
\end{ruledtabular}
\end{table}

\begin{equation}\label{Q0U}
\Delta U_0=\left(\frac{e}{2}-|Q_0|\right)\left(\frac{C_1+C_2}{C_1C_2}\right).
\end{equation}

Thereby, $Q_0$ can be related to the local work functions of the three metals according to \cite{Tinkham}:

\begin{equation}\label{Q0}
Q_0=\frac{1}{e}\left[C_1(\Phi_{\rm W}-\Phi_{\rm Co})-C_2(\Phi_{\rm Mo}-\Phi_{\rm Co})\right].
\end{equation}

The determined values for $Q_0$ are also summarized in table \ref{results}. They vary from cluster to cluster in the range of up to $\Delta Q_0=0.34 e$. The corresponding spatial variation of the Co work function can be calculated as $\Delta \Phi_{\rm Co}=eQ_0/(C_2-C_1)$, if one assumes that $\Phi_{\rm Mo}$ is spatially constant and that the tip did not change between individual measurements. The former is a good assumption, if the background charge is located close to the MgO-Co interface. The resulting values of $\Delta \Phi_{\rm Co}$ are added to table \ref{results}. Interestingly, the maximum change amounts to $\Delta\Phi_{\rm Co}=450$\,meV, which is rather close to the work function variation of the pure MgO film determined by FER spectroscopy. Thus, we assume that the variation of $\Delta U_0$ also results from positively charged color centers located close to the MgO surface. The slightly lower value of 450 meV probed by the Co clusters with respect to 600 meV measured on the pure MgO might be caused by screening of the color centers via Co islands, by a lateral distance between Co clusters and color centers and/or by averaging over the lateral size of the cluster.\\
Finally, we will discuss the full width at half maximum (FWHM) of the Coulomb peaks $\Delta E_{\rm P}$ determined after subtracting a linear background from the spectra in figure \ref{blockade}c)-f). The values of $\Delta E_{\rm P}$ averaged over the peaks of one curve are displayed in table \ref{results}. The peak width of about $300-400$\,meV is much larger than the energy resolution of the STS experiment, which is limited by the thermal energy ($T=300$\,K) and the applied modulation voltage ($U_{\rm mod}=40$\,mV) and given by $\Delta E_{\rm T}\approx\sqrt{(3.3\cdot kT)^2+(2.5\cdot eU_{\rm mod})^2}=130$\,meV \cite{Morgenstern}. Notice that this formula is in agreement with experiment down to $\Delta E_{\rm P}=0.1$\,meV accordingly $T=0.3$\,K \cite{Wiebe}. In order to cross-check, we varied the modulation voltage from 30\,mV to 70\,mV on a single Co island observing changes in the FWHM of less than 20 mV. Life time broadening of the peaks $\Delta E_L>h/2RC$ is orders of magnitudes smaller (in the range of tenth of $\mu$eV) and can also be neglected. Thus, the charge within the cluster is well equilibrated before leaving the cluster.
Similar FWHMs (270\,mV) have been observed previously for defect states of positively charged Cl vacancies in NaCl films even at $T=5$\,K \cite{Repp4}. They are attributed to an excitation of optical phonons in the NaCl lattice. Following this argument, we speculate that $\Delta E_{\rm P}$ is caused by the strong coupling of the tunneling electrons to optical phonons of the MgO-lattice. An estimate of the resulting FWHM is given by $\Delta E_{\rm P} =  \sqrt{8\ln{2}\cdot S}\cdot \hbar \omega$ with $S$ being the Huang-Rhys factor and $\omega$ being the relevant phonon frequency \cite{Repp4}. Using a typical optical phonon energy of MgO ($\hbar \omega \simeq 50$\,meV) \cite{Parlinski} and the Huang-Rhys factor $S=39$ determined for color centers in bulk MgO \cite{Henderson}, one gets $\Delta E_{\rm P} \simeq 700$\,meV, which is larger by a factor of two than the measured value. The larger value is plausible regarding the stronger electron-phonon coupling within the color center than the distant coupling taking place between electrons within the Co island and phonons within the MgO.
Alternatively, the observed width could also be due to charge fluctuations within the neighboring MgO, which are fast with respect to the measurement time, but slow on the time scale of a single electron tunneling event. These fluctuations can, of course also be induced by the tunneling current. However, in order to explain the observed width of $300-400$\,meV, single charges have to fluctuate within 2 nm of the cluster. We cannot exclude that enough defect states including chargeable cracks are close to each cluster. Thus, further studies are required in order to pinpoint the reason for the unexpectedly large FWHM.\\

\section{Conclusion}

In conclusion, we have studied the electronic properties of MgO films grown on Mo(100) as well as of Co nano-islands on top of them. Using field emission resonances observed in scanning tunneling spectroscopy at room temperature, we found fluctuations of the work function within a closed MgO film up to $\Delta \Phi=0.6$\,eV. These fluctuations could be attributed to positively charged defect states (maybe F+ color centers) close to the MgO surface. On the Co islands, we observed Coulomb oscillations at room temperature with energy distances in excellent agreement with orthodox theory. Analyzing the Coulomb gap we deduced a spatially varying background charge $Q_0$, which probably is also related to the charged color centers. The widths of the Coulomb peaks ($300-400$ meV) is considerably larger than the instrumental resolution. We speculate that a strong coupling between the electrons in the Co islands and the phonons of the MgO induces this increased width.

\section{acknowledgment}
We thank N. Nilius and J. Repp for helpful dicussions and G. G\"untherodt for the possibility to use his instrument.
We appreciate financial support of the German Science foundation via Mo 858/5-1. C. Pauly acknowledges additional funding by
"Fonds national de la recherche (Luxembourg)".

\end{document}